\documentclass[pdftex]{jps-cp}
\usepackage{txfonts} 
\usepackage{amsmath,amssymb,bm,braket,mathtools}
\usepackage[pdftex]{color}

\newcommand{\bket}[1]{\mathinner{|{#1}\rangle\!\rangle}}

\newcommand{\bbraket}[1]{\mathinner{\langle\!\langle{#1}\rangle\!\rangle}}

\title{Topological Phases in a PT-Symmetric Dissipative Kitaev Chain}

\author{Makio \textsc{Kawasaki}$^{1}$ and Hideaki \textsc{Obuse}$^{1,2}$}

\inst{$^{1}$Department of Applied Physics, Hokkaido University, Sapporo 060-8628, Japan\\
$^{2}$Institute of Industrial Science, The University of Tokyo, 5-1-5 Kashiwanoha, Kashiwa, Chiba 277-8574, Japan}

\email{makio{\_}k0620@eis.hokudai.ac.jp}

\recdate{July 15, 2022}

\abst{We study a topological phase in the dissipative Kitaev chain described by the Markovian quantum master equation.
Based on the correspondence between Lindbladians, which generate the dissipative time-evolution, and non-Hermitian matrices, Lindbladians are classified in terms of non-Hermitian topological phases.
We find out that the Lindbladian retains PT symmetry which is the prominent symmetry of open systems and then all the bulk modes can have a common lifetime.
Moreover, when open boundary conditions are imposed on the system, the edge modes which break PT symmetry emerge, and one of the edge modes has a zero eigenvalue.}

\kword{PT symmetry, Topological phase, Quantum master equation, Kitaev chain}

\begin{document}
\maketitle

\section{Introduction}
Parity-time (PT) symmetry is one of the most significant symmetries of open systems with gain and loss.
Various open systems can be described by non-Hermitian Schr\"{o}dinger equations.
If a non-Hermitian Hamiltonian and its eigenstates respect PT symmetry, the spectrum is entirely real \cite{1998Bender}.
PT-symmetric open systems can be realized in the classical \cite{2010Ruter} and quantum \cite{2017Xiao} optical setups and can be applied, for example, to the lasing \cite{2014Hodaei} and sensing \cite{2016Liu}.
Topological phenomena in PT-symmetric open systems are also intensively studied \cite{2015Yuce,2017Weimann,2018Kawabata,2020Kawasaki}.

However, the non-Hermitian Schr\"{o}dinger equation is an approximated time-evolution of the open quantum systems and can describe only the short-time dynamics.
Instead, the time evolution of density operators should be taken into account to describe the long-time dynamics of the open quantum systems.
If a system-environment coupling is sufficiently weak, the time-evolution of open quantum systems is well captured by the Markovian quantum master equation \cite{2002Breuer} \(\displaystyle i\frac{d\rho}{dt}=\hat{\mathcal{L}}[\rho]\).
Since the superoperator \(\hat{\mathcal{L}}\) called Lindbladian is non-Hermitian, we can introduce PT symmetry to the general open quantum systems \cite{2012Prosen}.
The topological phenomena of Markovian open quantum systems are studied from the dynamical perspective \cite{2020Lieu,2022Kawasaki} with the help of the non-Hermitian topological phases \cite{2019Kawabata}.
Nevertheless, the topological phases of PT-symmetric Markovian open quantum systems are still unclear.

In this work, we investigate the PT-symmetric topological phase of open quantum systems described by the Markovian quantum master equation.
We consider the Kitaev chain which is the one-dimensional topological superconductor with dissipation.
We show that the system respects PT symmetry and that all the bulk spectrum of the Lindbladian \(\hat{\mathcal{L}}\) has a common imaginary part.
We also show that the edge modes break PT symmetry, and one of them must have a zero eigenvalue in a wide parameter region.

\section{Dissipative topological superconductors with PT symmetry}
\subsection{Model and formalism} \label{subsec:model}
We consider a topological superconductor that couples with environments.
If the memory effects are negligible, the time-evolution is given as the Markovian quantum master equation with the Lindblad form \cite{2002Breuer},
\begin{equation}
  i\frac{d\rho}{dt}=\hat{\mathcal{L}}[\rho]=[\mathcal{H},\rho]+i\sum_{\mu}(2L_{\mu}\rho L_{\mu}^{\dagger}-\{L_{\mu}^{\dagger}L_{\mu},\rho\}), \label{eq:lindblad}
\end{equation}
where \(\rho\) is the density operator of the system.
\(\mathcal{H}\) is the Hamiltonian of the system, and the jump operator \(L_{\mu}\) describes a dissipation process of the system.
The superoperator \(\hat{\mathcal{L}}\) generates the time-evolution, and we call it Lindbladian.
In this work, we consider the Kitaev chain whose Hamiltonian is given as
\begin{equation}
  \mathcal{H}=\frac{it_0}{2}\sum_j(w_{j,\alpha}w_{j,\beta}-w_{j,\beta}w_{j,\alpha})+\frac{it_1}{2}\sum_j(w_{j,\beta}w_{j+1,\alpha}-w_{j+1,\alpha}w_{j,\beta}),~t_0,t_1\geq0 \label{eq:Kitaev}
\end{equation}
where \(w_{j,\alpha}\) and \(w_{j,\beta}\) are the Hermitian Majorana operators satisfying \(\{w_{j,s},w_{j',s'}\}=2\delta_{j,j'}\delta_{s,s'}\).
They are related to the fermionic operators \(a, a^{\dagger}\) as \(w_{j,\alpha}=a_j+a_j^{\dagger},~w_{j,\beta}=i(a_j^{\dagger}-a_j)\).
We focus on the one-body dissipation proportional to the Majorana operators:
\begin{equation}
  L_{j}=\gamma w_{j,\alpha},\quad\gamma>0. \label{eq:jump_op}
\end{equation}

Since the Hamiltonian is non-interacting and the jump operators are linear in the Majorana operators, the Lindbladian is also non-interacting and we can employ third quantization \cite{2008Prosen,2021Barthel}.
The set of operators of the \(n\) fermion systems form a Hilbelt space with the Hilbert-Schmidt inner product \(\bbraket{A|B}=\mathrm{tr}[A^{\dagger}B]\), and we can regard the operator \(A\) as a vector \(\bket{A}\).
We choose the basis as \(\bket{P_{\bm{p}}}=\bket{w_{1,\alpha}^{p_{1,\alpha}}w_{1,\beta}^{p_{1,\beta}}\cdots w_{n,\beta}^{p_{n,\beta}}}\) (\(p_{j,s}\in\{0,1\}\), and we use the notation \(\bm{o}\coloneqq(o_{1,\alpha},o_{1,\beta},\cdots,o_{n,\beta})\) where \(o\) is an arbitrary object such as number, operator, or superoperator, throughout the paper), then the fermionic superoperators
\begin{equation}
  \hat{c}_{j,s}\bket{P_{\bm{p}}}=\delta_{p_{j,s},1}\bket{w_{j,s} P_{\bm{p}}},~\hat{c}_{j,s}^{\dagger}\bket{P_{\bm{p}}}=\delta_{p_{j,s},0}\bket{w_{j,s} P_{\bm{p}}} \label{eq:3rdquant_basis}
\end{equation}
can be defined.
They satisfy the fermionic anticommutation relations \(\{\hat{c}_{j,s},\hat{c}_{j',s'}^{\dagger}\}=\delta_{j,j'}\delta_{s,s'},~\{\hat{c}_{j,s},\hat{c}_{j',s'}\}=0\).
Since the Lindbladian changes the basis \(\bket{P_{\bm{p}}}\) as \(\bket{w_{j,s} w_{j',s'} P_{\bm{p}}},~\bket{P_{\bm{p}}w_{j,s} w_{j',s'}},\) and \(\bket{w_{j,s} P_{\bm{p}} w_{j',s'}}\) by the non-interacting assumption,
we can rewrite the Lindbladian as a quadratic form of the fermionic superoperators \(\hat{c},~\hat{c}^{\dagger}\).

After some calculation, it is shown that the Lindbladian preserves the fermion parity \(\Pi=(-1)^{N}\) (\(N\) is the particle number \(N=\sum_j a_{j}^{\dagger} a_{j}\)) though the Lindbladian does not preserve the particle number \(N\).
Then the non-interacting Lindbladian can be block-diagonalized in even and odd fermion parity sectors.
We focus on the even-parity sector because proper quantum states have even fermion parity \(\Pi\rho\Pi=\rho\) (that is, they are written as the linear combination of the operators of the form \(\ket{\mathrm{even}}\bra{\mathrm{even}}\) and \(\ket{\mathrm{odd}}\bra{\mathrm{odd}}\), where \(\Pi\ket{\mathrm{even}}=\ket{\mathrm{even}}\) and \(\Pi\ket{\mathrm{odd}}=-\ket{\mathrm{odd}}\)).
We obtain a simplified form of the Lindbladian of this model for the even-parity sector as
\begin{equation}
  \hat{\mathcal{L}}=4\hat{\bm{c}}^{\dagger}Z\hat{\bm{c}}. \label{eq:3rdquant}
\end{equation}
The non-Hermitian matrix \(Z\) is defined by
\begin{equation}
  Z\coloneqq H-i\mathrm{Re}M, \label{eq:Z_def}
\end{equation}
where \(H\) and \(M\) are Hermitian matrices defined by the coefficients of the \(\mathcal{H}\) and \(L_{\mu}\):
\begin{gather}
  \mathcal{H}=\bm{w}^TH\bm{w},~M=\sum_{\mu}\bm{l}_{\mu}\bm{l}_{\mu}^{\dagger},~L_{\mu}=\bm{l}_{\mu}^{T}\bm{w}=\sum_{j,s} l_{\mu,j,s}w_{j,s}. \label{eq:M_def}
\end{gather}
Thus, we regard the Lindbladian as a non-Hermitian non-interacting Hamiltonian with the first quantized Hamiltonian \(4Z\).
By diagonalizing \(Z\), the Lindbladian is also diagonalized as
\begin{equation}
  \hat{\mathcal{L}}=\sum_{j=1}^{2n}4\lambda_j \hat{b}'_j\hat{b}_j, \label{eq:lind_diag}
\end{equation}
where \(Z=\sum_{j=1}^{2n}\lambda_j \bm{\psi_j}\bm{\chi_j}^{\dagger}\) is the eigendecomposition of \(Z\) and \(\hat{b}'_j\coloneqq \hat{\bm{c}}^{\dagger}\bm{\psi_j},~\hat{b}_j\coloneqq \bm{\chi_j}^{\dagger}\hat{\bm{c}}\) are creation and annihilation superoperators of the eigenmodes satisfying generalized canonical anticommutation relations \(\{\hat{b}_j,\hat{b}'_k\}=\delta_{j,k},~\{\hat{b}_j,\hat{b}_k\}=\{\hat{b}'_j,\hat{b}'_k\}=0\).
Non-interacting Lindbladians always have a steady state \(\bket{\mathrm{NESS}}\) such that \(\hat{b}_j\bket{\mathrm{NESS}}=0\) for all \(j\).
Then an eigenoperator of the Lindbladian is constructed as \(\prod_{j=1}^{2n}\hat{b}'_j{}^{\nu_j}\bket{\mathrm{NESS}}~(\nu_j\in\{0,1\})\), and its eigenvalue is given as \(4\sum_{j=1}^{2n}\lambda_j\nu_j\).
In particular, an eigenoperator with even fermion parity and a zero eigenvalue corresponds to a steady state of the system \cite{2021Barthel}.
Under the periodic boundary conditions, \(Z\) in our model is given in the momentum space as
\begin{equation}
  Z(k)=\frac{i}{2}\begin{pmatrix}
  -2\gamma^2 & -t(k)^* \\ t(k) & 0
\end{pmatrix},~t(k)\coloneqq t_0+t_1 e^{ik}. \label{eq:Z_kit}
\end{equation}

\subsection{PT symmetry and the topological invariant}
Owing to Eq.\ \eqref{eq:3rdquant}, we can investigate some properties of the Lindbladian by investigating the matrix \(Z\) instead.
In particular, we investigate the topological properties of the dissipative Kitaev chain in this work.
To this end, we employ the topological classification of the non-Hermitian matrix \(Z\) \cite{2019Kawabata,2020Lieu,2022Kawasaki}.
\(Z\) has all symmetries in AZ\(^{\dagger}\) class, and it belongs to class BDI\(^{\dagger}\):
\begin{align}
  \mathrm{TRS}^{\dagger}:~&\mathcal{T} Z^T(-k)\mathcal{T}^{-1}=Z(k),~\mathcal{T}=\sigma_z, \label{eq:trs} \\
  \mathrm{PHS}^{\dagger}:~&Z^*(-k)=-Z(k), \label{eq:phs} \\
  \mathrm{CS}:~&\Gamma Z^{\dagger}(k)\Gamma^{-1}=-Z(k),~\Gamma=\sigma_z, \label{eq:cs}
\end{align}
where \(\sigma_z\) is one of the Pauli matrices \(\sigma_z=\begin{pmatrix}1&0 \\ 0&-1\end{pmatrix}\).
Moreover, the traceless part of \(Z\) has PT symmetry as
\begin{equation}
  (\mathcal{PT})\left[Z(k)+\frac{i\gamma^2}{2}I\right]^*(\mathcal{PT})^{-1}=Z(k)+\frac{i\gamma^2}{2}I,~\mathcal{PT}=\sigma_x, \label{eq:pt}
\end{equation}
where \(\sigma_x=\begin{pmatrix}0&1\\1&0\end{pmatrix}\) and \(I\) is the identity operator.
Note that some symmetries such as PT symmetry for Lindbladians are defined for the traceless part \cite{2012Prosen,2022Kawasaki} as the Lindbladians do not have eigenstates with the positive imaginary part of eigenvalues.
PT symmetry in Eq.\ \eqref{eq:pt} ensures that the spectrum of \(Z\) are of the form \(\nu-i\gamma^2/2~(\nu\in\mathbb{R})\) or \(\{\nu-i\gamma^2/2,\nu^*-i\gamma^2/2\}~(\nu\in\mathbb{C})\).
The eigenvalues take the former if the corresponding eigenvectors of \(Z\), \(\bm{\psi}\), is the same with \(\mathcal{PTK}\bm{\psi}\) up to phase factors,
where \(\mathcal{K}\) is the complex conjugation operation.
On the other hand, the eigenvalues take the latter if $\bm{\psi}$ and \(\mathcal{PTK}\bm{\psi}\) are linearly independent.
We say that the eigenvectors do not break PT symmetry if the corresponding eigenvalues take the former, while 
the eigenvectors break PT symmetry and forms a pair \(\{\bm{\psi},\mathcal{PTK}\bm{\psi}\}\) if the corresponding eigenvalues take the latter.
Typically, eigenvalues in the former form become the latter form as we increase the non-Hermiticity of a PT-symmetric matrix.
If \(Z\) does not have PT symmetry breaking eigenvectors, in other words, imaginary parts of all eigenvalues are \(-i\gamma^2/2\), the system belongs to the PT symmetry unbroken phase.
If some eigenvectors break PT symmetry, {\it i.e.} imaginary parts of some eigenvalues are different from \(-i\gamma^2/2\), the system belongs to the PT symmetry broken phase.
We call the transition from PT unbroken phase to broken phase PT symmetry breaking.

\begin{figure}[tb]
  \centering
  \includegraphics[width=0.95\columnwidth]{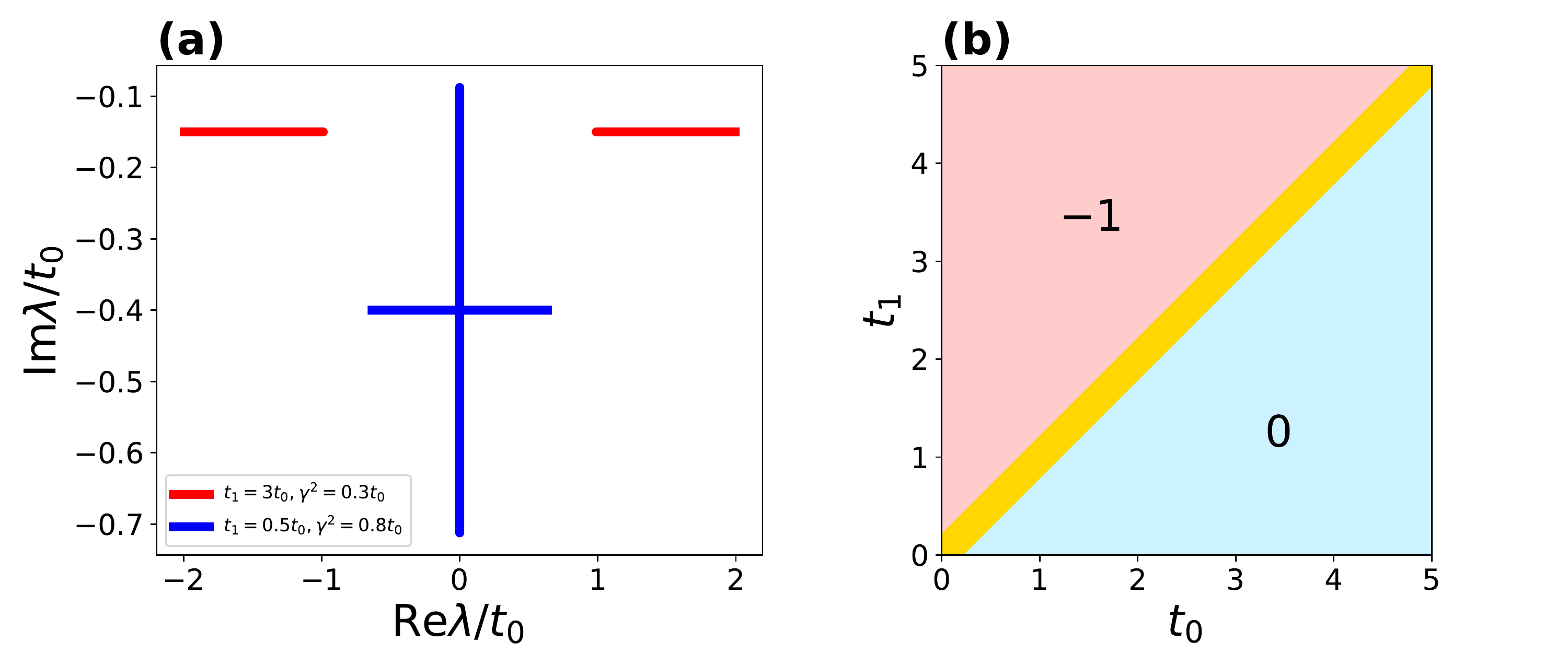}
  \caption{(a)The spectrum of \(Z\) with periodic boundary conditions.
  Red and blue lines correspond to the cases of \(t_1=3t_0,\gamma^2=0.3t_0\) (PT symmetry unbroken) and \(t_1=0.5t_0,\gamma^2=0.8t_0\) (PT symmetry broken), respectively.
  (b)The phase diagram of the topological invariant in Eq.\ \eqref{eq:winding} when \(\gamma^2=0.2t_0\).
  In the yellow region, the real line gap closes due to PT symmetry breaking, and the winding number in Eq.\ (\ref{eq:winding}) is ill-defined.}
  \label{fig:winding}
\end{figure}

The eigenvalues of \(Z(k)\) in Eq.\ \eqref{eq:Z_kit} are obtained as
\begin{equation}
  \lambda_{\pm}=\pm\frac{1}{2}\sqrt{|t(k)|^2-\gamma^4}-\frac{i\gamma^2}{2}. \label{eq:dispersion}
\end{equation}
We show the eigenvalues in Fig.\ \ref{fig:winding} (a).
The imaginary parts of all eigenvalues equal to \(-i\gamma^2/2\) if \(|t(k)|^2>\gamma^4\) for all \(k\) [red lines of Fig.\ \ref{fig:winding} (a)].
All eigenvectors do not break PT symmetry, and the real line gap opens in this region.
The PT symmetry breaking occurs and the line gap closes if there exists \(k\) such that \(|t(k)|^2\leq\gamma^4\) [blue vertical line of Fig.\ \ref{fig:winding} (b)].
The left and right eigenvectors of $Z(k)$ are written as
\begin{equation}
  \bm{\chi_{\pm}}^{\dagger}=\frac{1}{N_{\pm}}(\pm\sqrt{|t(k)|^2-\gamma^4}-i\gamma^2,-it(k)^*),~\bm{\phi_{\pm}}=\frac{1}{N_{\pm}}\binom{\pm\sqrt{|t(k)|^2-\gamma^4}-i\gamma^2}{it(k)}, \label{eq:eigenvector}
\end{equation}
where \(N_{\pm}\) is the normalization constant as \(N_{\pm}^2=2\sqrt{|t(k)|^2-\gamma^4}(\sqrt{|t(k)|^2-\gamma^4}\mp i\gamma^2)\).

Now we focus on the topological properties of the Lindbladian \(\hat{\mathcal{L}}\) in Eq.\ (\ref{eq:3rdquant}).
According to the topological classification of non-Hermitian matrices \cite{2019Kawabata}, the matrices in class BDI\(^{\dagger}\) with one spatial dimension have a \(\mathbb{Z}\)-valued topological invariant.
We can calculate a topological invariant as the winding number,
\begin{equation}
  w=\frac{1}{2\pi i}\int_{\mathrm{BZ}}q^{-1}\frac{dq}{dk}dk, \label{eq:winding}
\end{equation}
where \(q\) is defined through the matrix \(Q\):
\begin{equation}
  Q=\begin{pmatrix}0&q \\ q^{\dagger}&0\end{pmatrix},~Q\coloneqq I-(\bm{\phi_-}\bm{\chi_-}^{\dagger}+\bm{\chi_-}\bm{\phi_-}^{\dagger}). \label{eq:q_def}
\end{equation}
Inserting the eigenvectors into \(Q\), we get that the winding number is 0 if \(t_0>t_1+\gamma^2\) and -1 if \(t_1>t_0+\gamma^2\).
The phase diagram of the topological invariant is shown in Fig.\ \ref{fig:winding} (b).

\subsection{PT symmetry breaking edge modes}
\begin{figure}[tb]
  \centering
  \includegraphics[width=0.95\columnwidth]{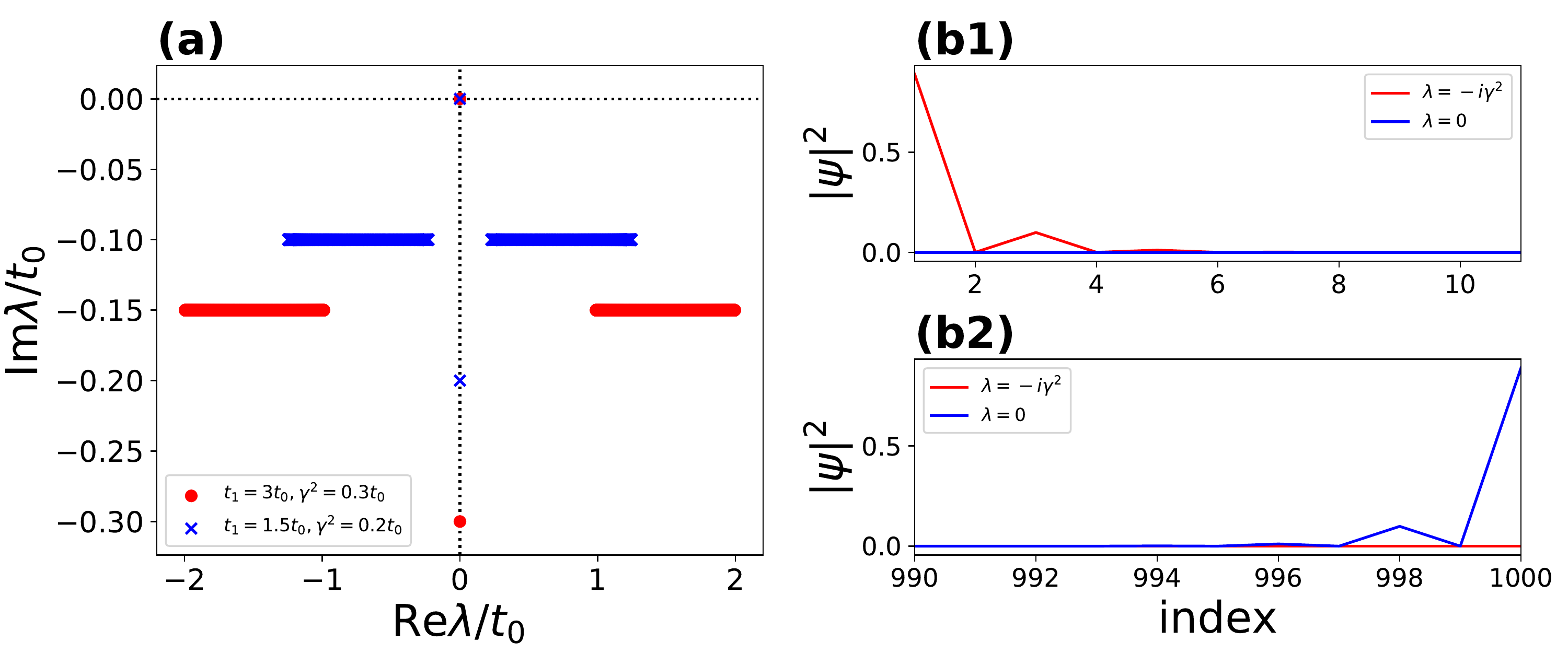}
  \caption{(a) The spectra of \(Z\) with open boundary conditions.
  The red dots and blue crosses correspond to the cases of \(t_1=3t_0,\gamma^2=0.3t_0\) and \(t_1=1.5t_0,\gamma^2=0.2t_0\), respectively.
  (b) Spatial configuration of edge modes in the case of \(t_1=3t_0,\gamma^2=0.3t_0\).
  We set \(n=500\).
  If the index is odd (even), the flavor of the corresponding position is \(\alpha~(\beta)\).
  (b1) Edge states near the left boundary.
  (b2) Edge states near the right boundary.}
  \label{fig:eig}
\end{figure}

We study the topological edge modes of the Lindbladian in Eq.\ \eqref{eq:3rdquant} in this subsection.
We numerically diagonalize \(Z\) for open boundary conditions in the parameter region so that all bulk modes do not break PT symmetry, and the spectra are shown in Fig.\ \ref{fig:eig}.
We confirmed that two edge modes appear in Fig.\ \ref{fig:eig} (a), as predicted by the bulk-edge correspondence with the topological invariant we have calculated in the previous subsection.
All bulk modes do not break PT symmetry, and corresponding eigenvalues have the same imaginary part \(-i\gamma^2/2\), as shown in Fig. \ref{fig:eig} (a).
In contrast, the two edge modes break PT symmetry since the edge modes localize near the one boundaries as shown in Fig.\ \ref{fig:eig} (b).
Their eigenvalues takes \(0\) and \(-i\gamma^2\) in both cases.
We can show the existence of these edge modes analytically.
If the eigenvector of \(Z\) is written as \((A_1,B_1,\cdots,A_n,B_n)^T\), the eigenvalue equation of \(Z\) is recast as
\begin{align}
  -t_1B_{j-1}-2\gamma^2 A_j+t_0B_j &= \frac{2}{i}\lambda A_j, \label{eq:eigen_bulk1} \\
  -t_0A_j+t_1A_{j+1} &= \frac{2}{i}\lambda B_j \label{eq:eigen_bulk2}
\end{align}
for the bulk and
\begin{align}
  -2\gamma^2 A_1+t_0B_1 &= \frac{2}{i}\lambda A_1, \label{eq:eigen_edge1} \\
  -t_0A_n &= \frac{2}{i}\lambda B_n \label{eq:eigen_edge2}
\end{align}
for the boundaries.
We note that the above equations correspond to the eigenvalue equation of the Kitaev chain $\mathcal{H}$ in Eq.\ (\ref{eq:Kitaev}) when \(\gamma=0\).
If \(t_1>t_0\), a zero-energy edge mode of the Kitaev chain $\mathcal{H}$ satisfying the above equations in the limit of $n \rightarrow \infty$ with \(A_j=0,~B_j/B_{j-1}=t_1/t_0\) is also the solution of the corresponding eigenvalue equation of \(Z\) with \(\lambda=0\).
The edge mode localizes near $j=n$ as shown in the blue curve of Fig.\ \ref{fig:eig} (b2).
Since the dissipation acts only on the flavor \(\alpha\), the edge mode is unaffected by the dissipation and the eigenvalue remains at zero.
Conversely, the other zero-energy edge mode of the Kitaev chain $\mathcal{H}$ satisfying Eqs.\ \eqref{eq:eigen_bulk1} - \eqref{eq:eigen_edge2} in the limit of $n \rightarrow \infty$ with \(A_j/A_{j+1}=t_1/t_0,~B_j=0\) is most sensitive to the dissipation, which leads to the largest imaginary part of the eigenvalue; \(\lambda=-i\gamma^2\).
The latter edge mode localizes near the other boundary $j=1$ as shown in the red curve of Fig.\ \ref{fig:eig} (b1).
The latter edge mode with \(\lambda=-i\gamma^2\) is also obtained by applying \(\mathcal{PTK}\) to the former edge mode because they form a pair due to PT symmetry breaking.
We also confirm that the edge modes do not appear in the parameter region with \(w=0\), as predicted by the bulk-edge correspondence.

Finally, we mention the steady state of the system.
Since the jump operators in this model are Hermitian, the infinite-temperature state \(\rho_{\mathrm{inf}}\propto I\) is the steady state of the time-evolution \(\hat{\mathcal{L}}[\rho_{\mathrm{inf}}]=0\).
Although \(Z\) has a zero eigenvalue, it does not lead to multiple steady states (see also Refs.\ \cite{2019Caspel,2021Barthel}).
We recall that the spectrum of \(\hat{\mathcal{L}}\) is expressed by the spectrum of \(Z\) as \(4\sum_{j=1}^{2n}\lambda_j\nu_j\) and the eigenoperators are given as \(\prod_{j=1}^{2n}\hat{b}'_j{}^{\nu_j}\bket{\mathrm{NESS}}\).
Since \(\hat{b}'\) changes the fermion parity, the eigenoperators with \(\sum_j\nu_j\) even have even fermion parity and correspond to the quantum states.
In particular, from the formulae, degenerated zero eigenvalues of \(Z\) are essential to have additional steady states of the system.

\section{Conclusions}
We have studied the topological phase of the dissipative Kitaev chain in this work.
We have shown that the dissipative Kitaev chain retains PT symmetry and the bulk spectrum of the Lindbladian can possess a common imaginary part due to PT symmetry.
Imposing open boundary conditions on the system, we have clarified that the edge modes break PT symmetry and have different imaginary parts of the eigenvalues from the bulk modes.
We have discussed that the eigenvalues of edge modes must be \(0\) and \(-i\gamma^2\), while the steady edge mode is impossible in this model.
The future direction of this work is the systematic construction of steady edge states by utilizing PT symmetry.
This work sheds light on manipulating the steady topological edge states by engineering dissipation.

We thank Yasuhiro\ Asano and Kousuke\ Yakubo for helpful discussions.
M.\ K.\ was supported by JST SPRING (Grant No.\ JPMJSP2119).
This work was also supported by KAKENHI (Grants No.\ 20H01828, No.\ JP21H01005, and No.\ JP22H01140, and No.\ 22K03463).

\end{document}